\documentclass[12pt]{article}
\usepackage{graphicx}

\newcommand\pubnumber{Article 26 in eConf C1304143}
\newcommand\pubdate{\today}

\def\napoli{Center for Cosmology and Particle Physics, Physics Department\\
 New York University, New York, NY 10003}

\def\Title#1{\begin{center} {\Large #1 } \end{center}}
\def\Author#1{\begin{center}{ \sc #1} \end{center}}
\def\Address#1{\begin{center}{ \it #1} \end{center}}

\newcommand\pubblock{\rightline{\begin{tabular}{l} \pubnumber\\
         \pubdate  \end{tabular}}}
\newenvironment{Abstract}{\begin{quotation}  }{\end{quotation}}
\newenvironment{Presented}{\begin{quotation} \begin{center} 
             PRESENTED AT\end{center}\bigskip 
      \begin{center}\begin{large}}{\end{large}\end{center} \end{quotation}}
\def\Acknowledgements{\bigskip  \bigskip \begin{center} \begin{large}
             \bf ACKNOWLEDGEMENTS \end{large}\end{center}}

\def\beq{\begin{equation}}
\def\eeq#1{\label{#1}\end{equation}}
\def\eeqn{\end{equation}}

\def\beqa{\begin{eqnarray}}
\def\eeqa#1{\label{#1}\end{eqnarray}}
\def\eeqan{\end{eqnarray}}

\let\bar=\overbar

\def\Dslash{\not{\hbox{\kern-4pt $D$}}}
\def\dslash{\not{\hbox{\kern-2pt $\del$}}}

\def\msb{{\bar{\ssstyle M \kern -1pt S}}}

\begin{document}
\begin{titlepage}
\pubblock

\vfill
\Title{GRB Afterglow Blast Wave Encountering Sudden Circumburst Density Change Produces No Flares}
\vfill
\Author{Ilana Gat, Hendrik van Eerten, Andrew MacFadyen}
\Address{\napoli}
\vfill
\begin{Abstract}
Afterglows of gamma-ray bursts are observed to produce light curves with the flux following power law evolution in time. However, recent observations reveal bright flares at times on the order of minutes to days. One proposed explanation for these flares is the interaction of a relativistic blast wave with a circumburst density transition. In this paper, we model this type of interaction computationally in one and two dimensions, using a relativistic hydrodynamics code with adaptive mesh refinement called \textsc{ram}, and analytically in one dimension. We simulate a blast wave traveling in a stellar wind environment that encounters a sudden change in density, followed by a homogeneous medium, and compute the observed radiation using a synchrotron model. We show that flares are not observable for an encounter with a sudden density increase, such as a wind termination shock, nor for an encounter with a sudden density decrease. Furthermore, by extending our analysis to two dimensions, we are able to resolve the spreading, collimation, and edge effects of the blast wave as it encounters the change in circumburst medium. In all cases considered in this paper, we find that a flare will not be observed for any of the density changes studied.
\end{Abstract}
\vfill
\begin{Presented}
Huntsville Gamma Ray Burst Symposium\\
Nashville, Tn, USA,  April 14--18, 2013
\end{Presented}
\vfill
\end{titlepage}
\def\thefootnote{\fnsymbol{footnote}}
\setcounter{footnote}{0}

\section{Introduction}\label{sec:intro}


Gamma-ray burst (GRB) afterglows are generally modeled as smooth curves with the slope being a function of the density of the surrounding medium as well as the power-law slope of the distribution of the accelerated electron population at the shock front. Recent observations, though, have shown flares in X-ray afterglows. To explain the causes of these flares, researchers have began to study the interaction of a blast wave with complex structures such as a wind termination shock (e.g., \cite{Ramirez-Ruiz2005,Peer2006, Nakar2007}). In this paper, we revisit this interaction by modeling the circumburst density structure as a stellar wind with a sudden density increase followed by a homogeneous medium (e.g., an interstellar medium (ISM)). We also extend this analysis to density drops as well.


We use a numerical relativistic hydrodynamics (RHD) code with adaptive mesh refinement called \textsc{ram} \cite{Zhang2006} to numerically simulate these interactions in one and two dimensions. Our simulations are set up to address a number of interaction scenarios not previously explored as well as the two dimensional effects of spreading of a collimated flow at the shock front. Specifically, we address the following questions in our analysis:
\begin{itemize}
\item When a collimated blast wave traveling in a stellar wind environment encounters a wind termination shock, how does the size of the density jump affect the dynamics and resulting light curve?
\item What happens when there is a density drop instead of a jump? Does the blast wave speed up, and in turn, recollimate? Will this cause a rebrightening or flare in the light curve?
\item When the blast wave encounters an extreme density increase, does it immediately spread outwards from this high energy collision and cause flares in the light curve? Does this sideways spreading depend on the size of the density jump?
\end{itemize}

We answer these questions through our numerical and analytical simulations. These results are explained in detail in~\cite{Gat2013}, which can also be referenced for a more complete list of relevant citations.

\section{Dynamics of Blast Wave Encounters}\label{sec:D}

\begin{figure*}[htb]
\centering
\includegraphics[width=0.49\textwidth]{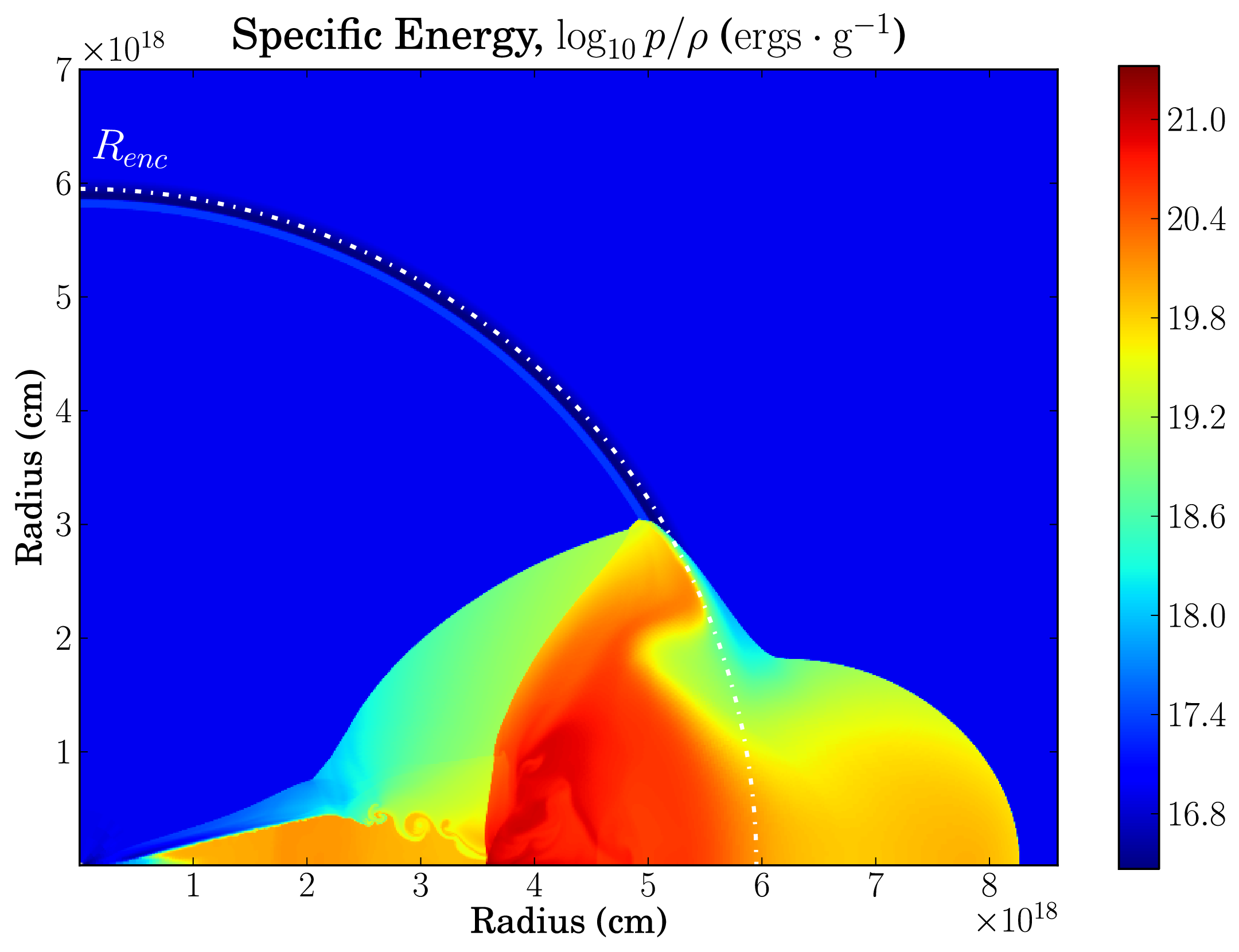}
\includegraphics[width=0.49\textwidth]{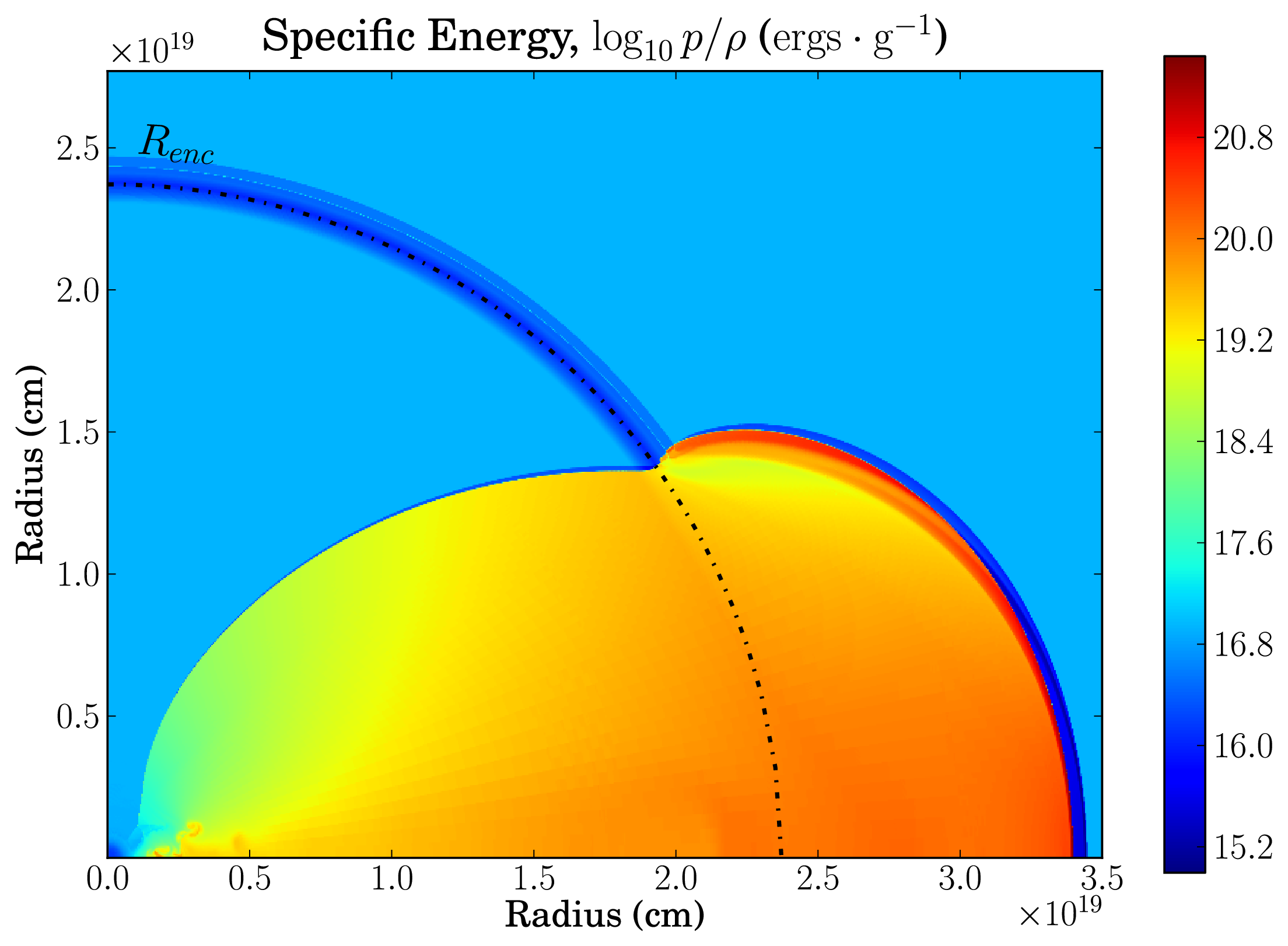}
\caption{Figures of the two dimensional numerical simulations showing specific energy. Left: simulation of a blast wave encountering a density jump of factor 100 at a radius corresponding to the fluid shock Lorentz factor, $\gamma=10$ at a time, $T=2.97\times10^8s$. Right: simulation of a blast wave encountering a density drop of factor 100 at a radius corresponding to the fluid shock Lorentz factor, $\gamma=5$ at a time, $T=1.15\times10^9s$. }
\label{fig:2d_se}
\end{figure*}

Our numerical simulations of the adiabatic blast wave (``jet'') formed by a GRB are set up in spherical coordinates following the Blanford \& Mckee solution~\cite{Blandford1976}. They have a jet half opening angle of $\theta_0 = 0.1$ and a starting fluid Lorentz factor, $\gamma=15$. Figure~\ref{fig:2d_se} shows two snapshots: one from a simulation of a blast wave encountering a density jump (left panel) and the other from a simulation of a blast wave encountering a density drop (right panel). The left panel in Figure~\ref{fig:2d_se} shows the recollimation of the blast wave as it enters the higher density region as well as the strong reverse shock caused from the encounter. Conversely, the right panel in Figure~\ref{fig:2d_se} illustrates the spreading of the blast wave as it encounters the much lower density region. These results, at first, seem counter-intuitive as a blast wave traveling with low Lorentz factors usually spreads and a blast wave with high Lorentz factors is more collimated. Here, the blast wave that is slowed by the higher density region recollimates, and the blast wave that speeds up spreads out. This is explained by the fact that the blast wave traveling into a higher density medium has more pressure pushing back at it, forcing it to recollimate as it pushes into the new region. The blast wave traveling into the lower density region has much less pressure against it, allowing the blast wave to spread as it travels into the ISM.

In addition to the recollimation of the blast wave encountering a large density jump (left panel in Figure~\ref{fig:2d_se}), the reverse shock created from the encounter shocks and spreads the fluid still in the stellar wind environment. This causes vortices to form that get trapped within the stellar wind environment and never pass through to the ISM. There is also a small amount of sideways spreading from the actual encounter itself, however this is a very small amount. The most energetic fluid punctures through to the higher density medium at the time of the encounter instead of spreading outwards. Yet, sideways spreading is still observed as the reverse shock spreads the fluid remaining in the stellar wind environment. The blast wave encountering a density drop (right panel in Figure~\ref{fig:2d_se}) does not create a strong reverse shock or spread sideways in the stellar wind, and thus the fluid behind the drop is still able to travel through to the new medium.

With the understanding of the dynamics of the blast wave encountering a sudden change in circumburst medium, the next section discusses the resulting light curves.

\section{Light Curves}\label{sec:LC}

We calculate the GRB afterglow light curves using the radiation calculation methodology from~\cite{vanEerten2010offaxis} at an X-ray frequency of $5\times10^{17}$ Hz, similar to the frequency detected by {\it Swift}. We studied light curves at other frequencies, however the results were qualitatively the same.  We use values of $\epsilon_B = 0.01$, $\epsilon_e = 0.1$, and $p = 2.5$, where $\epsilon_B$ and $\epsilon_e$ are the fractions of internal energy that contribute to the magnetic field at the shock front and to accelerating electrons respectively, and $p$ expresses the energy distribution index of the shock-accelerated particles. We also do not consider electron cooling in our light curves shown here, but have found that this does not qualitatively change our results.

\begin{figure*}[!htb]
\centering
\includegraphics[width=0.49\textwidth]{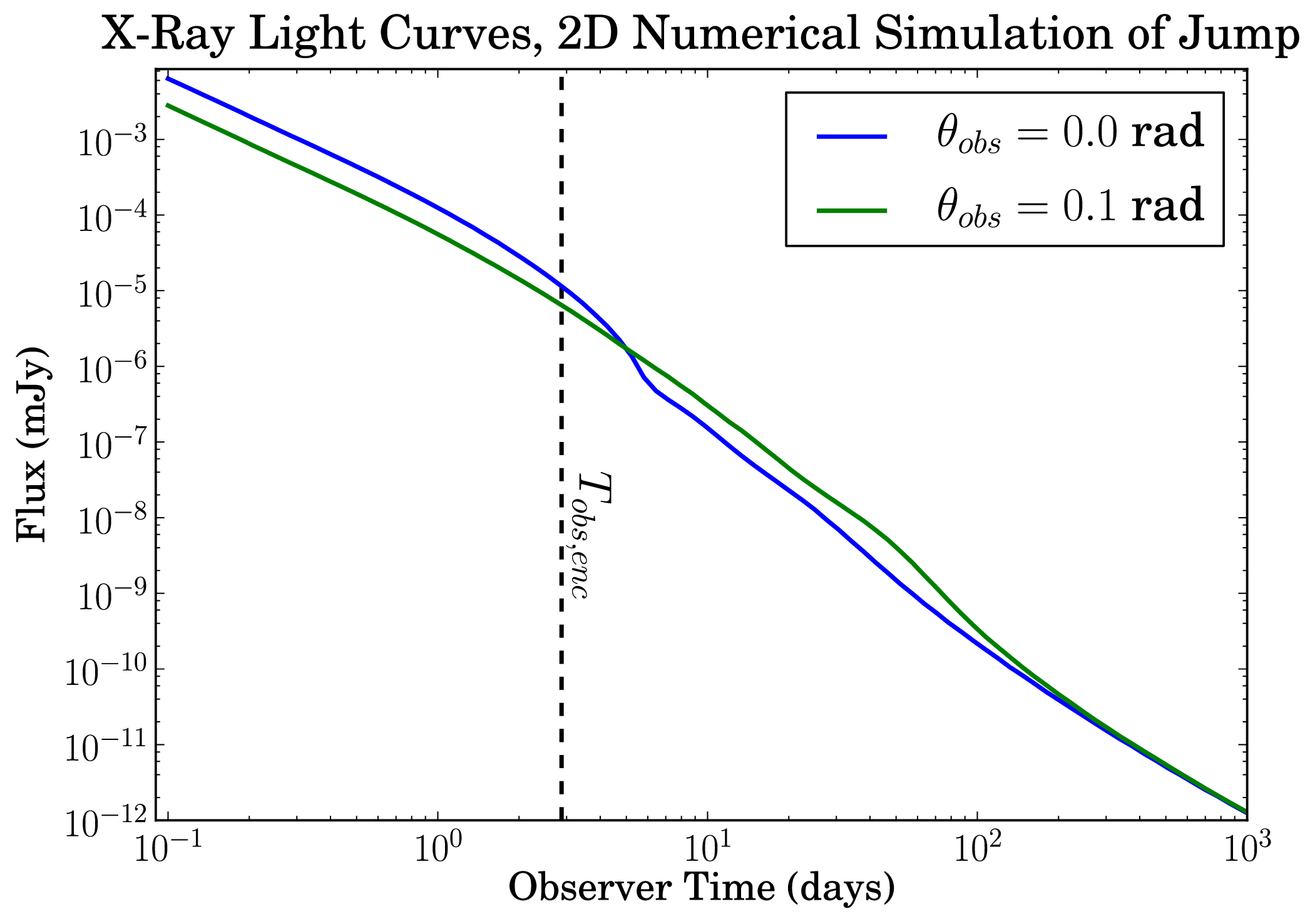}
\includegraphics[width=0.49\textwidth]{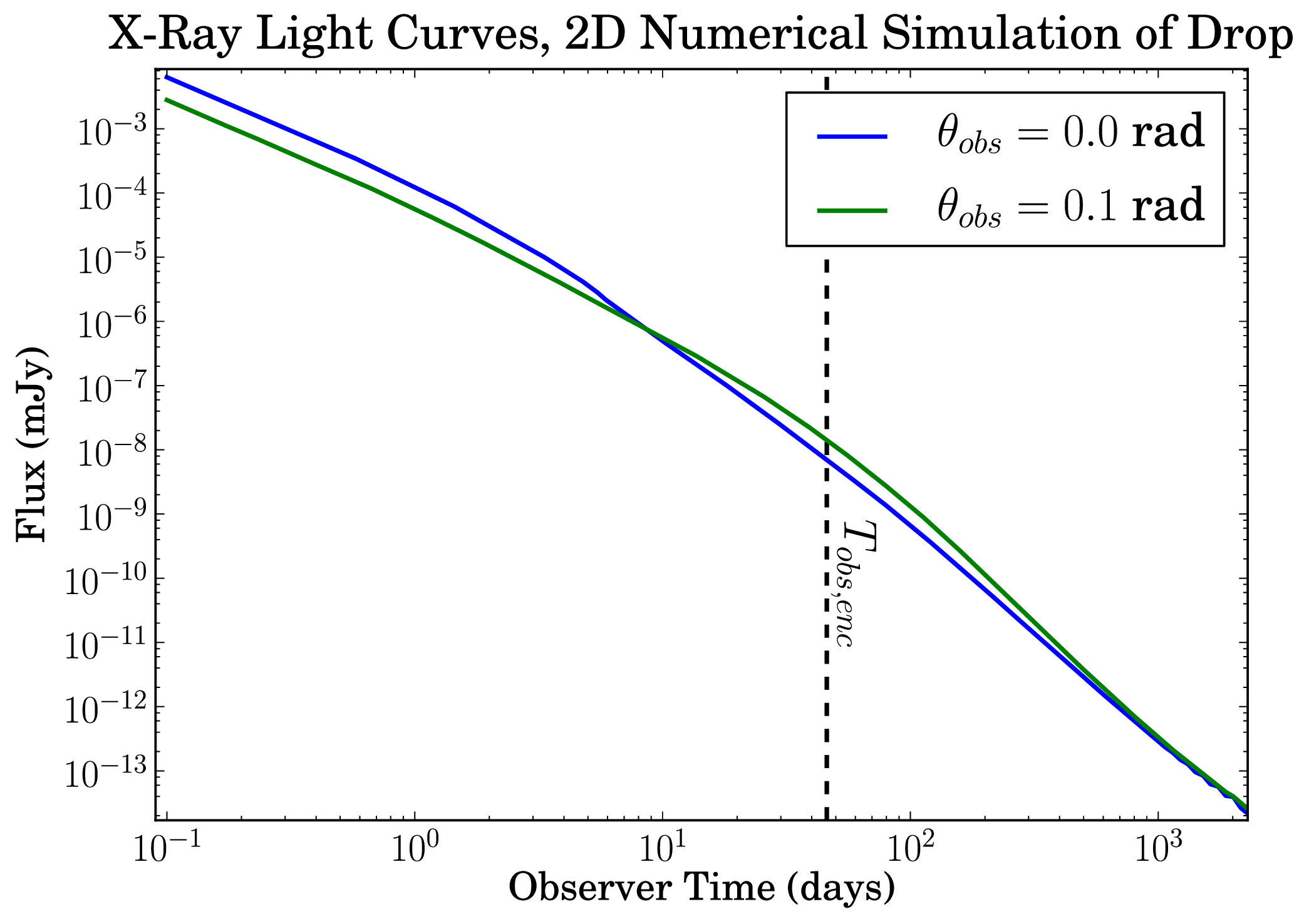}
\includegraphics[width=0.49\textwidth]{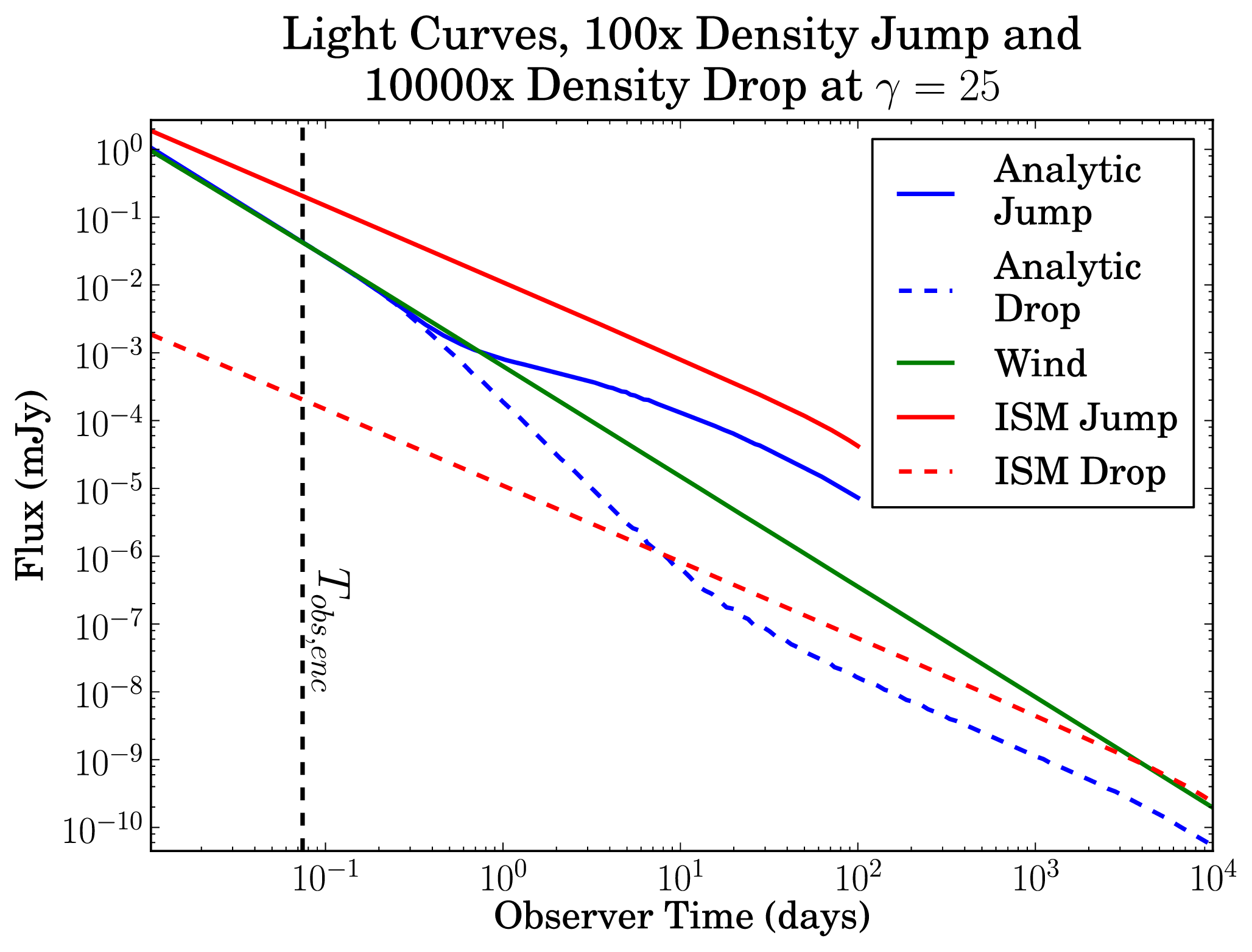}
\includegraphics[width=0.49\textwidth]{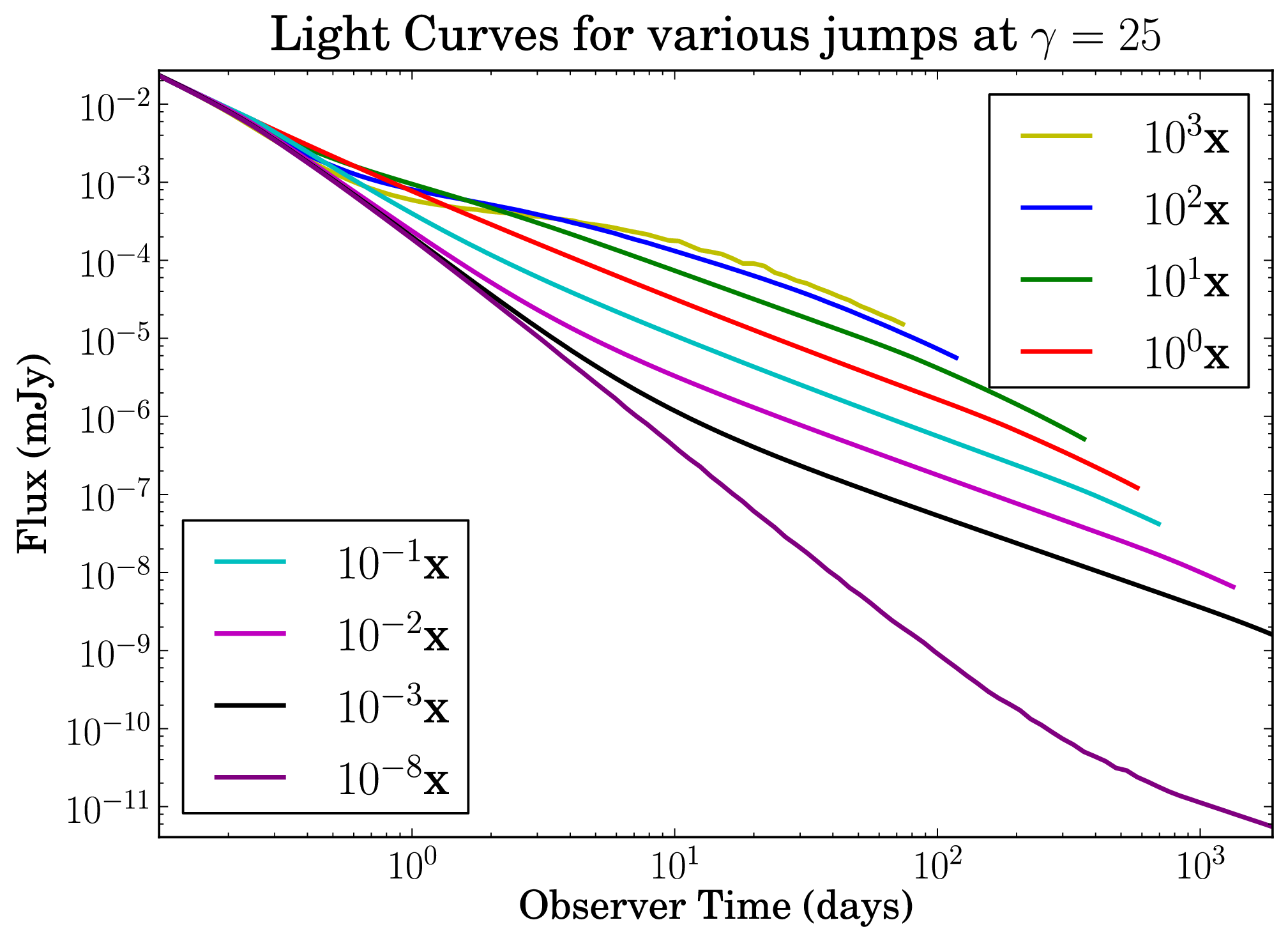}
\caption{Top: light curves from the two dimensional numerical simulations.Top Left: blast wave encountering a density jump of factor 100  at a radius corresponding to the fluid shock Lorentz factor, $\gamma=10$ for on and off axis observations. Top Right: blast wave encountering a density drop of factor 100  at a radius corresponding to the fluid shock Lorentz factor, $\gamma=5$ for on and off axis observations. Bottom: light curves from the analytical model (described in~\cite{Gat2013}) assuming spherical outflow for a blast wave encountering a density change at a radius corresponding to the fluid shock Lorentz factor, $\gamma=25$. Bottom Left: blue light curves are of blast wave traveling in a stellar wind environment encountering a density jump of factor 100 (solid blue line) and a density drop of factor 10000 (dashed blue line) followed by a homogeneous medium. These are plotted against the light curve for a blast wave traveling solely in a stellar wind environment (green line) and the light curves for blast waves traveling solely in the homogeneous medium of the jump (red solid line) and of the drop (dashed solid line). Bottom Right: light curves for a blast wave encountering encountering a density change of various magnitudes.}
\label{fig:lcs}
\end{figure*}

Figure~\ref{fig:lcs} depicts that there are no observable flares in the flux emitted from a blast wave encountering a jump or a drop. The time labeled ``$T_{obs,enc}$'' in the plots of Figure~\ref{fig:lcs} is the first time the encounter should be observable. The two dimensional simulations of a density jump (top left panel in Figure~\ref{fig:lcs}) show that there will be no flare at the time of the encounter for an observer on or off axis. However, the magnitude of the light curve will drop and the slope of the light curve will change. For the case of the density drop, the light curve also smoothly transitions to a new slope after the encounter, which can be seen in the top right panel of Figure~\ref{fig:lcs}. 

When a blast wave traveling in a stellar wind encounters a change in density followed by an ISM, the slope of the flux observed will change after the encounter to match the slope of the flux observed if the blast wave was solely traveling in the ISM. This is depicted in the bottom left panel in Figure~\ref{fig:lcs}. This smooth transition to the new slope results in no sudden increase in flux. For the case of a density jump, the flux does not immediately jump up to the new magnitude because the blast wave has been slowed by the encounter and is not energetic enough to cause a rebrightening. For the case of a drop, the flux observed from a blast wave in the new ISM environment is at a lower magnitude than the flux observed from a blast wave in the wind environment. This results in the flux observed from the blast wave encountering the new environment to decrease, instead of increase, after the encounter to evolve in the new environment. For both cases, though, of a drop or a jump, the magnitude of the flux observed after the encounter is lower than the flux would have been if the blast wave was traveling solely in the ISM. This is due to the different regions of the post-encounter blast wave. The fluid swept up prior to the encounter is held behind the contact discontinuity and all the fluid swept up after the encounter is contained in front of the contact discontinuity. Thus, there is much less fluid radiating at the post encounter Lorentz factor than there would be if the blast wave was traveling solely in the ISM, resulting in the magnitude of the flux observed to be lowered.

The bottom light panel in Figure~\ref{fig:lcs} shows the light curves from a blast wave traveling in a stellar wind environment that encounters jumps of various sizes at a radius corresponding to the fluid Lorentz factor, $\gamma=25$. This figure illustrates that the conclusion that the flux observed will smoothly transition to the slope of the light curve in the new medium holds for a wide range of jump factors. At the time of the encounter, the flux observed from the blast wave will not suddenly increase. It will transition to the new base line in the new medium.

\section{Conclusions}\label{sec:Con}

We have shown numerically and analytically that a blast wave evolving partially in a stellar wind environment that encounters a sudden change in density, either an increase or a decrease, followed by a constant density environment for a wide range of initial conditions does not cause an observable rebrightening. The size of the density jump does affect the dynamics and resulting light curve, but there are still no observable flares at the time of the encounter. 

We found that for a blast wave traveling in a stellar wind environment encountering an ISM environment, the resulting flux observed will gradually transition from one environment to the next. If the flux observed from a blast wave traveling solely in the ISM is lower than the flux observed with a blast wave traveling solely in the wind environment at the time of the encounter, the flux observed from the blast wave will simply dim and follow the same light curve slope of the ISM. If the flux observed from a blast wave traveling solely in the ISM is higher than the flux observed from a blast wave traveling solely in a wind environment at the time of the encounter, the observed flux from the blast wave will not suddenly increase, but stay relatively steady as it transitions to the new slope of the ISM.

We have studied the two dimensional and one dimensional effects of a density drop and have shown that the blast wave does increase in speed, but does not recollimate. There is also no rebrightening caused by this sudden increase in blast wave speed. Our two dimensional studies of a density jump have yielded the conclusion that there is some sideways spreading from a high energy collision of the blast wave with a large jump in circumburst density, and the amount by which the blast wave spreads is highly dependent on the size of the density jump. 

From our analytical solutions and our numerical simulations, we have answered the questions listed in the Introduction and have concluded that a blast wave traveling in a stellar wind environment that encounters a change in density followed by an ISM environment will not cause observable flares. We conclude that a wind termination shock, or more generally, any sudden transition in circumburst density (even extreme changes), is very unlikely to be the cause of the flares observed.  In addition, flares observed by {\it Swift} return to the same baseline as the light curve prior to the flare \cite{burrows2005}. If the flare occurred at the time of the encounter, the light curve would transition to the new slope of the ISM material, not the slope of the stellar wind. This makes it highly unlikely that a flare seen at the encounter with the change in circumburst environment could be the explanation for the flares seen by {\it Swift}.

\Acknowledgements
This research was supported in part by NASA through grant NNX10AF62G issued through the Astrophysics Theory Program and by the NSF through grant AST-1009863 and by the Chandra grant TM3-14005X. Resources supporting this work were provided by the NASA High-End Computing (HEC) Program through the NASA Advanced Supercomputing (NAS) Division at Ames Research Center. The software used in this work was in part developed by the DOE-supported ASCI/Alliance Center for Astrophysical Thermonuclear Flashes at the University of Chicago.\\


\begin{thebibliography}{99}

\bibitem{Ramirez-Ruiz2005}
  E.~Ramirez-Ruiz {\it et al.},
  Astrophys.\ J.\  {\bf 631}, 435 (2005)


\bibitem{Peer2006}
  A.~Pe'er and R.~A.~M.~J.~Wijers,
  Astrophys.\ J.\  {\bf 643} 1036 (2006) 


\bibitem{Nakar2007}
  E.~Nakar and J.~Granot,
  Mon.\ Not.\ Roy.\ Astron.\ Soc.\  {\bf 380}, 1744 (2007)


%

\bibitem{Zhang2006}
W.~Zhang and A.~I.~MacFadyen,
 Astrophys.\ J.\ Suppl. {\bf 164}, 255 (2006)


\bibitem{Gat2013} 
  I.~Gat, H.~van Eerten and A.~MacFadyen, 
  Astrophys.\ J.\  {\bf 773} 2 (2013)

\bibitem{Blandford1976}
  R.~D.~Blandford and C.~F.~McKee,
  Phys.\ Fluids {\bf 19}, 1130 (1976).


\bibitem{vanEerten2010offaxis}
  H.~van Eerten, W.~Zhang and A.~MacFadyen,
  Astrophys.\ J.\  {\bf 722}, 235 (2010)

\bibitem{burrows2005} 
 D.~N.~Burrows {\it et al.},
  Science {\bf 309}, 1833 (2005)


\end{thebibliography}
\end{document}